\documentclass[12pt]{article}

\newcommand{\be}{\begin{equation}}
\newcommand{\ee}{\end{equation}}
\newcommand{\bea}{\begin{eqnarray}}
\newcommand{\eea}{\end{eqnarray}}
\newcommand{\nn}{\nonumber}
\newcommand{\cD}{\mathcal{D}}
\newcommand{\lb}{\left[}
\newcommand{\rb}{\right]}
\newcommand{\half}{\frac{1}{2}}
\newcommand{\gP}{\bar\mathcal{P}}
\newcommand{\Pg}{\mathcal{P}}
\newcommand{\gh}{\rm gh}
\newcommand{\bQ}{\bar Q}
\newcommand{\ra}{\rightarrow}

\begin{document}
\begin{titlepage}

\title{Duality between constraints and gauge conditions }

\author{M. Stoilov\\
   {\it Institute of Nuclear Research and Nuclear Energy,
        Sofia 1784, Bulgaria}\\
e-mail: {\rm mstoilov@inrne.bas.bg}}

\date{10 May 2005}
\maketitle

%\keywords{gauge theory, BRST charge}
%\preprint{0504220}

\begin{abstract}
It is shown that in the first order gauge theories under some general assumptions
gauge conditions can play the role of new local symmetry generators,
while the original constraints become gauge fixing terms.
It is possible to associate with this new symmetry a second BRST charge and its 
anticommutator with the original BRST charge is the Hodge operator of the 
corresponding cohomology complex.
\end{abstract}

\thispagestyle{empty}
\end{titlepage}

%\thispagestyle{empty}
%\setcounter{page}{0}
%\pagebreak

Gauge theories are of a permanent interest for theoretical physics.
They are best understood in the Hamiltonian approach to systems 
with constraints proposed by Dirac \cite{D}. 
In the Hamiltonian approach one does not work in the configuration space 
of the model but in the larger phase space which includes
besides the dynamical coordinates also their momenta.
As a result of this enlargement of the number of variables, there is
a huge freedom in the choice of the coordinates and momenta
describing one and the same physics with the equivalent sets of variables
connected by the so called ``canonical transformations''.
The simplest example of a canonical transformation
is the mutual interchange of a coordinate $q$ and
its momentum $p$: $q\ra p,\;\;p\ra -q$.
This transformation is fundamental for the understanding of
the gauge condition -- constraint duality.
The reason is that, loosely speaking, in a particular coordinate 
system in the phase space the gauge conditions are part of the dynamical
coordinates and the constraints are the corresponding momenta, both
forming the unphysical sector of the theory.
Performing the mentioned above canonical transformation in this unphysical
sector, one in fact interchanges the constraints and the gauge conditions. 
Therefore, one can view the gauge fixed model as originated from a
different gauge theory in which the local symmetry is
generated by the gauge conditions of the initial model and the
former constraints now play the role of gauge conditions.
In this dual picture we can change the gauge conditions
(former constraints) because they are now a matter of our choice.
Thus, performing the gauge conditions-constraints duality twice we
can produce a theory quite different from the initial one.

Let us introduce some notations.
In what follows we consider a gauge theory with Hamiltonian $H$
and with constraints $\varphi_a\;\;a=1,\dots,m$,  all of them
functions of the phase space variables $q$ and $p$.
For simplicity we suppose that $\varphi$ are first class Bose constraints
and that the entire model is of order one.
Together these two requirements mean that the following Poisson bracket 
relations hold:
\bea
 \lb\varphi_a, \varphi_b\rb & = &C_{abc}\varphi_c, \label{alg1}\\
\lb H, \varphi_a\rb & = & U_{ab}\varphi_b, \label{al}
\eea
where $U_{ab}$ and $C_{abc}$ do not depend on dynamical variables.
A proper treatment of such model requires supplementary gauge 
fixing conditions $\chi_a$. 
They could be arbitrary functions of $q$ and $p$ but have 
to form an Abelian algebra  \cite{FS}
\be
\lb\chi_a,\chi_b\rb=0\;\;\;\forall a,b.\label{ab}
\ee 
With their help it is possible to write down the transition amplitude
(or S-matrix) for the theory in the functional integral representation
\be
Z = \int \cD p \cD q 
exp\left\{ i\int_t \left(p\dot q - H(p,q)\right){\rm d t}\right\} 
\prod_a\delta(\varphi_a)\delta(\chi_a) 
{\rm det}\vert \Delta\vert. \label{gf}
\ee
Here 
\be
\Delta = \lb\chi_b,\varphi_c\rb \label{del}
\ee
must be invertible and an implicit summation over all degrees of freedom
(which could be discrete as well as continuance) is understood.

It is clear that if we consider a new gauge model in which the
new constraints are the former gauge conditions $\chi_a$ and 
the new gauge conditions are the former constraints $\varphi_a$
it will possess the same transition amplitude as (\ref{gf}) and
will describe the same physics.
However, there are some self-consistency
conditions which constraints and gauge conditions should obey.
One of them is related to eqs.(\ref{ab}) which 
at first sight seems to be crucial. 
These conditions are a consequence of the requirement that after a suitable
canonical transformation $\chi_a$  become part of the new 
coordinates (in fact --- the first $m$ ones).
Then it is possible to  resolves the constraints $\varphi_a=0$ 
with respect to the corresponding to $\chi_a$ momenta 
(which produce the determinant term in (\ref{gf})). 

Our primary task here is to show that eqs.(\ref{ab}) could be relaxed.
It is enough for our purposes to assume that $\chi_a$ form instead of
(\ref{ab}) the following non--Abelian algebra 
\be
\lb\chi_a, \chi_b\rb  = D_{abc}\chi_c \label{nab}
\ee
with $D_{abc}$ independent of the dynamical variables.
The easiest way to show that this assumption does not change the
transition amplitude (\ref{gf}) is to introduce notations allowing  
symmetric treatment of both $\varphi_a$ and $\chi_a$. 
Let $\phi_a$ denote the set of constraints and gauge conditions
\be
\phi_{\bf a}=\{\varphi_a, \chi_a\}. 
\ee
Now we have a model with (second class
\footnote{The fact that $\phi_{\bf a}$ are second class constraints allows
to use the results of ref.\cite{EM} for a direct solution of the problem.}) 
constraints
\be
\phi_{\bf a}=0 \label{f1}
\ee
and we want to find the physical degrees of freedom in it.
This is possible only if eqs.(\ref{f1}) are solvable for 
(the first $m$) canonical pairs
$\zeta_{\bf a}=\{ q_a, p_a\}$ (which we suppose), or equivalently,
the following unequality should be fulfilled:
\be
det\vert\frac{\partial\phi_{\bf a}}{\partial\zeta_{\bf b}}\vert\ne 0.
\label{f2} 
\ee
Eq.(\ref{f2}) allows us to write down the transition amplitude for the 
considered model and it reads
\be
Z = \int \cD p \cD q 
exp\left\{i \int_t \left(p\dot q - H(p,q)\right){\rm d t}\right\} 
\prod_{\bf a}\delta(\phi_{\bf a})
{\rm det}\vert\frac{\partial\phi_{\bf a}}{\partial\zeta_{\bf b}}\vert. 
\label{gff}
\ee
Constraints (\ref{f1}) determine physical submanifold in the entire 
phase space.
The normals to this submanifold correspond to the unphysical degrees of freedom
 $\zeta_{\bf a}$.
Using eqs.(\ref{f1}) and (\ref{f2}) these variables can be determined  
 through the physical ones.
Note that $\zeta_{\bf a}$ are not uniquely determined --- any local 
$sp(m)$ transformation gives another set of unphysical variables 
$\zeta_{\bf a}'$ which is as good as the initial one.
We use this freedom to set 
\be
\frac{\partial\chi_a}{\partial p_b}=0.\label{f3}
\ee
The physical coordinates  (which we denote by $\zeta^*$) 
should be  complementary to $\zeta_{\bf a}$, i.e., they should
describe the tangential to the surface (\ref{f1}) directions and we have
\be
\frac{\partial\phi_{\bf a}}{\partial\zeta^*}=0. \label{f4}
\ee
Using eqs.(\ref{f3},\ref{f4}) it is easy to see that
\be
{\rm det}\vert\frac{\partial\phi_{\bf a}}{\partial\zeta_{\bf b}}\vert =
{\rm det}\vert\left[\chi_a,\varphi_b\right]\vert,
\ee
and so, the transition amplitude (\ref{gff}) coincides with (\ref{gf}).
This proves that it is possible to use non-Abelian gauge conditions provided 
${\rm det}\vert \left[\chi_b,\varphi_c\right]\vert\ne 0$.

Another selfconsistency condition which should be fulfilled
concerns the Poisson brackets between $H$ and $\chi_a$.
We assume that equations, analogous to eqs.(\ref{al}) hold
\be
\lb H, \chi_a\rb = V_{ab}\chi_b \label{alg2}
\ee
with $V_{ab}$ independent of the phase space variables.
Eqs.(\ref{al}),(\ref{alg2}) impose severe restrictions on the
Hamiltonian form.
The only compatible with them expression is
\be
H=\varphi_a F_{ab} \chi_b + H_{ind} \label{ham}
\ee
where $H_{ind}$ commutes both with $\varphi_a$ and $\chi_a$ for each $a$.
For the matrices $U$ and $V$ we get
\be
U^T = \Delta F = - V. \label{mat}
\ee
Note that when $\varphi_a$ and/or $\chi_a$ form non-Abelian algebra,
$\Delta$ depends on the dynamical variables and so does $F$.
There are some other relations between $F$ and $C$ and between $F$ and $D$
which are consequence of eq.(\ref{ham}) and the first order requirement
which we shall not use.

It is known that the gauge fixed action is BRST invariant.
The BRST charge is constructed entirely on bases of the gauge symmetry
algebra (\ref{alg1}) \cite{Hen}. 
Our aim here is to show that there is a duality between constraints
and gauge conditions.
So we want to construct a second BRST charge, determined by
the algebra (\ref{nab}) of the constraints.
 
We begin with writing down the BRST charge. 
Here we are using a modification of the Batalin -- Fradkin -- Vikovisky
BRST charge.
This modified BRST charge depends on the Hamiltonian of the model 
in consideration  \cite{m}.
For its construction we need two systems of ghost--antighost pairs 
$\{c_a,\gP_a\}$ and $\{\bar c_a,\Pg_a\}$,
$\lb c_a, \gP_b\rb = -\delta_{ab} = \lb\bar c_a, \Pg_b\rb$ and
with the opposite ghost numbers
$
\gh(c_a) = - \gh(\gP_a) = \gh(\Pg_a) = -\gh(\bar c_a) = 1.
$
The BRST charge reads:
\be
Q=c_a(\varphi_a - U_{ab} \pi_b + C_{abc}\lambda_b\pi_c + 
C_{abc}\Pg_b\bar c_c) + 
\half c_a c_b C_{abc}\gP_c  + i\Pg_a \pi_a. 
\label{b2}
\ee
It differs from the BRST charge in the BFV approach  \cite{Hen}
by the term $c_a(- U_{ab} \pi_b + C_{abc}\lambda_b\pi_c + 
C_{abc}\Pg_b\bar c_c)$.
This term, together with $i\Pg_a \pi_a$ originates from such alteration
of the constraints so that now they
generate the gauge transformation of the Lagrange multipliers too.

According to the BFV procedure the BRST invariant action of the model is
\be
S=\int  \dot q p + \dot\lambda \pi + \dot c \gP + \dot{\bar c}\Pg -
H' + \lb Q,\psi\rb, \label{act}
\ee
where $\psi$ is an arbitrary imaginary anticommuting function 
and $H'$ contains beside the initial Hamiltonian $H$
also some ghost terms (see eq.(\ref{h}) below).
A basic choice for $\psi$ in the BFV approach is
\be
\psi=i\bar c_a \chi_a + \gP_a \lambda_a, \label{g3}
\ee
where it is supposed that $\chi_a$ are functions of $q$ and $p$ only
which obey eqs.(\ref{ab}). 
In the spirit of our previous considerations, it is natural to recognize
in eq.(\ref{g3}) after a canonical change of variables $\lambda\ra\pi$,
$\pi\ra -\lambda$ the (multiplied by $i$) BFV BRST charge for the
Abelian ``constraints'' $\chi_a$.
When $\chi_a$ form the non--Abelian algebra (\ref{nab})
we use the arbitrariness of $\psi$ to replace it by $\bQ$ ---
a newly introduced second BRST charge corresponding to the 
gauge symmetry, generated by $\chi_a$ 
\be
\bQ=\bar c_a(\chi_a - V_{ab} \lambda_b - D_{abc}\pi_b\lambda_c +
D_{abc}\gP_b c_c) +
\half \bar c_a \bar c_b D_{abc}\Pg_c  - i\gP_a \lambda_a. \label{b3}
\ee

Before proceeding further we need to refine the Hamiltonian $H'$
in (\ref{act}).
Initially it was constructed to be $Q$ invariant.
Now we want also $\bQ$ invariance and we expect that a
modification of the Hamiltonian ghost part will be needed.
The Hamiltonian we start with is  \cite{m}
\be
H' = H +\lambda_a U_{ab} \pi_b + c_a U_{ab}\gP_b
+ \Pg_a U_{ab} \bar c_b. \label{h}
\ee 
The last three terms ensure the $Q$ invariance of $H'$
(for the BRST charge defined by eq.(\ref{b2})). 
It turns out that, as a consequence of the eqs.(\ref{mat}),
this Hamiltonian is also $\bQ$ invariant, so $H'$ from (\ref{h})
is the Hamiltonian we shall work with.

Substituting all our formulas in eq.(\ref{act}) we obtain
a  simple expression for the double--BRST invariant action
\be 
S=\int \dot q p + \dot\lambda \pi + \dot c \gP + \dot{\bar c}\Pg -
H-\lambda\varphi + \pi\chi + i\bar c \Delta c + \lambda U \pi
 -i\chi C c \bar c -i\varphi D \bar c c + i\gP\Pg
+ L_{ext}, \label{brsta}
\ee
where $L_{ext}$ is a ghost term which we separate for a 
reason which we shall comment later.
Note that $\dot\lambda\pi$ term can be attached either to
$\lambda\varphi$ or to $\pi\chi$ thus producing the `Lorenz' gauge
for dual theories.

Using the action (\ref{brsta}) we can obtain the corresponding 
transition amplitude
as a functional integral of $e^{i S}$ over all phase space variables
(matter coordinates, Lagrange multipliers, ghosts and their momenta).
We want to compare thus obtained expression with the one
given by eq.(\ref{gf}).
In order to do this we exploit an idea of ref.\cite{Hen} to
perform a rescaling of the gauge conditions $\chi_a$ plus a change of 
variables whose Berezian is equal to $1$.
However in our case we need to rescale also the constraints $\varphi_a$.
All together our manipulations look as follow:
\bea
\chi_a \ra \frac{1}{\beta}\chi_a;&\pi_a=\beta\pi_a',& 
\bar c_a =\beta\bar c_a', \nn\\
\varphi_a \ra \frac{1}{\alpha}\varphi_a;&\lambda_a=\alpha\lambda_a',& 
c_a = \alpha c_a'. \label{res}
\eea
Note that as a consequence of eqs.(\ref{res}) the structure constants
as well as $\Delta$ are modified
\be
D_{abc}\ra\frac{1}{\beta}D_{abc}; \;\;
C_{abc}\ra\frac{1}{\alpha}C_{abc}; \;\;
V_{ab}\ra V_{ab};\;\;\Delta_{ab}\ra\frac{1}{\alpha\beta}\Delta_{ab}.
\ee
We consider the limit $\alpha\ra\infty ,\;\;\beta\ra\infty$ in which
$L_{ext},\;\;\dot\lambda \pi,\;\;\dot c \gP$ and $\dot{\bar c}\Pg$ 
go to zero and the transition amplitude for the 
model with action (\ref{brsta}) takes the form (for clearness we 
omit the sign $'$ in some of the variable notations)
\be
\bar Z = \int \cD(p q\pi\lambda\gP\Pg c\bar c)
exp\left\{ i\int \dot q p - 
H + i\bar c \Delta c + \pi\chi - i\chi C c \bar c
-\lambda\varphi -i\varphi D \bar c c + i\gP\Pg \right\}. \label{gf2}
\ee
The integral over momenta $\gP$ and $\Pg$ 
is trivial giving an overall normalization constant.
The term $ i\chi_a C_{abc} c_b \bar c_c$ can  
be absorbed in $\pi_a\chi_a$
by redefinition of $\pi_a$ and the same is possible for 
$i\varphi_{a} D_{abc} \bar c_b c_c$ which can be absorbed in 
$\lambda_a\varphi_a$.
After that integrations over $\lambda,\;\;\pi,\;\;c$ and $\bar c$ are
easily performed giving
\be
\bar Z = Z.
\ee

Finally, we want to make some comments about cohomological character of
our considerations \cite{FL}.
The functions over the full phase space formed by the dynamical variables, 
Lagrange multipliers and their momenta and the two systems of ghosts 
is an associative supercommutative algebra $\mathcal F$. 
This algebra has a natural grading with respect to the ghost number operator. 
The action of the BRST charge $Q$ on $\mathcal F$ 
gives to this superalgebra the structure of a graded
differential algebra.
Here $Q$, which has ghost number $1$, plays the same  role 
as the operator of the exterior derivative $d$
in the case of differential forms.
$\bQ$ which has ghost number $-1$ plays the role of $d^*$ --- the 
Hodge dual to $d$.
The term $i \lb Q, \bQ\rb$ which replaces the term $\lb Q,\psi\rb$
in the action (\ref{act}) is, in fact, the Hodge operator for $\mathcal F$.

%=============================================================================
\section*{Acknowledgements}
%=============================================================================

This work is supported by the
Bulgarian National Science Foundation, Grant Ph-1010/00.

\vfill\pagebreak

\end {document}